\documentclass[smallextended]{svjour3}       
\smartqed  
\usepackage{graphicx}
\usepackage{amssymb,exscale,epsfig,multirow}
\usepackage{pgf,pgfarrows,pgfnodes,pgfautomata,pgfheaps,pgfshade}
\usepackage{amsmath,amssymb}
\usepackage{graphicx}
\usepackage{enumerate}
\usepackage[latin1]{inputenc}
\usepackage{colortbl}
\usepackage{color}
\usepackage{tcolorbox}
\usepackage[english]{babel}

\begin{document}

\title{Pseudoscalar meson dominance and nucleon structure\thanks{
Supported by  PID2020-114767GBI00
funded by MCIN/AEI/10.13039/501100011033, as well as Junta de
Andaluc\'{\i}a grant FQM-225}
}


\author{E. Ruiz Arriola  \and P. S\'anchez Puertas 
}


\institute{Enrique Ruiz
  Arriola  and Pablo Sanchez Puertas \at
         Departamento de
  F\'{\i}sica At\'omica, Molecular y Nuclear and Instituto Carlos I
  de F{\'\i}sica Te\'orica y Computacional. Universidad de Granada,
  E-18071 Granada, Spain. \\ 
              \email{earriola@ugr.es}        and          \email{pablosanchez@ugr.es}           
}

\date{Received: date / Accepted: date}

\maketitle

\begin{abstract}
  Pseudoscalar meson dominance has implications for  nucleon structure
  which follow from an Extended Partial Conservation of the Axial Current
  (EPCAC). The minimal
  resonance saturation of the nucleon pseudoscalar form factor of the
  lowest pseudoscalar and isovector mesons compatible with pQCD short
  distance constraints and chiral symmetry. Using PDG
  masses and widths we obtain $g_{\pi NN} = 13.21(^{+0.11}_{-0.06})$,
  to be compared with the most precise determinations from $np, pp$
  scattering, $g_{\pi^+ np} = 13.25(5)$ from the Granada-2013
  database. Equivalently a Goldberger-Treiman discrepancy $\Delta_{\rm
    GT} = 1.8^{+0.9}_{-0.4}\%$ is found.  Our results are
  consistent with almost flat strong pion-nucleon-nucleon vertices.
\end{abstract}

\section{Introduction}

The ancient Goldberger-Treiman (GT) relation~\cite{Goldberger:1958tr}
is the most spectacular prediction of chiral symmetry where weak and
strong interaction properties are intertwined. The GT discrepancy is defined 
as
\begin{eqnarray}
  \Delta_{\rm GT}=1-\frac{M_N g_A}{g_{\pi^+ p n} f_{\pi^+}}   \, , 
\end{eqnarray}
where $M_N=(M_p+M_n)/2$ is the nucleon mass, $f_{\pi^+}$ the pion weak
decay constant, $g_A$ the axial nucleon coupling constant and
$g_{\pi^+ p n}$ the pion nucleon coupling constant. Using up to date
values $M_p=0.93827231$GeV and $M_n=0.93956563$ GeV, $g_A=1.2723(23)$
and $f_{\pi^+}=0.09206(84)$GeV from
PDG~\cite{ParticleDataGroup:2022pth} and $g_{\pi^+ pn}=
13.25(5)$~\cite{Perez:2016aol,Arriola:2016hfi} we get $\Delta_{\rm
  GT}= 2(1) \%$, i.e.  a small but significantly non-vanishing
discrepancy. The uncertainty obtained by standard error propagation in
quadratures~\footnote{ This means, under the assumption of uncorrelated uncertainties the relation $(\delta \Delta_{\rm
    GT})^2/\Delta_{\rm GT}^2 = (\delta f_{\pi^+})^2/ f_{\pi^+}^2+
  (\delta g_A)^2/ g_A^2+
  (\delta M_N)^2/M_N^2+ (\delta g_{\pi^+ p n})^2 /g_{\pi^+ p n}^2 $.  } is
dominated by the $0.2\%$ bench-marking accuracy of $g_{\pi^+ pn}$ from
a partial wave analysis of the about $np, pp$ existing 8000 elastice
scattering data below pion production threshold and reduced down to
6713 by the Granada-2013 $3\sigma$-selfconsistent database with
$\chi^2 /{\rm DOF}=1.025$. We review here our recent
work~\cite{ERA-PSP:2023} which is summarized in an almost back of the
envelope calculation with rather similar central value and uncertainty
exploiting pseudoscalar meson dominance.

At the hadronic level the GT relation is 
a direct consequence of Partial Conservation of the Axial Current (PCAC)~\cite{Nambu:1960xd} and the fact that in the isospin limit the
vector weak current or the neutron $\beta$-decay corresponds to the
isospin rotated strong conserved current~\cite{Feynman:1958ty}, (for a
reviews covering up to the mid 80's see
e.g.~\cite{Pagels:1974se,Dominguez:1984ka} and references therein).
It is remarkable that such a pre-QCD relation works so well with modern data.

Since the discovery of QCD, the GT relation is regarded as an exact
theorem in the chiral limit (massless quarks). For quark fields $q_i$
and $q_f$ PCAC in QCD reads
\begin{eqnarray}
\partial_\mu ( \bar q_f \gamma^\mu \gamma_5 q_i) = (m_f+m_i) \bar q_f i \gamma_5 q_i \
{-\frac{\alpha_s}{8\pi}\epsilon^{\mu\nu\rho\sigma} G^a_{\mu\nu}G^a_{\rho\sigma} \delta_{fi}  }  \, ,
\end{eqnarray}
where the last term, corresponding to the $U_A(1)$ anomaly, 
is ignored in the following as we shall deal with non-singlet currents. 
The first term on 
the right hand side is called the pseudoscalar density which has
$J^{PC}=0^{-+}$ quantum numbers. Assuming isospin symmetry, i.e.
$m_u=m_d \equiv m_0 $, the pseudoscalar density for light $u,d$
flavors, $2 m_0 \bar q \vec \tau i \gamma_5 q $ has $I^G=1^-$ quantum
numbers and therefore at the hadronic level it has a non vanishing
overlap with any state with an odd number of pions, $\pi, 3 \pi,
5\pi, \dots$ and the vacuum.

\section{Pseudoscalar form factor of the Nucleon}

Matrix elements between nucleon states
with initial and final momenta $p$ and $p'$ correspond to the nucleon
{\it pseudoscalar} form factor,
\begin{eqnarray}
  \langle N(p') | \bar q \{ \vec \tau , \hat m \} i \gamma_5  q  |N(p)  \rangle = \bar u(p') \{ \vec \tau , \hat m \} i \gamma_5 u(p) F_P (q^2) \, , 
\end{eqnarray}
where $q_{\mu}=p'_{\mu}-p_{\mu}$ is the momentum transfer and
$N=(p,n)$ $u(p)$ are nucleon Dirac spinors. The form factor satisfies
useful analytical properties in the complex $t-$plane.
\begin{enumerate}
\item $F_P(t)$ is real in the space-like region, $t < 0$, and up to the opening
  of the $3\pi$ threshold $t < (3
  m_\pi)^2 $,
\item 
  It has a pion pole at $t=m_\pi^2$ ,
\item  $F_P(t)$ has a
branch cuts along the odd number of pions production thresholds,
$t=(3m_\pi)^2,(5 m_\pi)^2 , \dots$, corresponding to the process $N \bar N \to 3 \pi, 5 \pi , \dots $
\item  $F_P(t)$  falls off as $ m_q (\alpha
(Q^2)/Q^2)^2$ in the deep Euclidean region, $t=-Q^2 \to - \infty $~\cite{Alvegard:1979ui,Brodsky:1980sx}
\item  Its value at the origin is $ F_P(0)= 2 M g_A $.
\end{enumerate}
With all these
properties being fulfilled we can write the following dispersion relation
\begin{eqnarray}
 \hat m F_P (t) = \frac{2 f_\pi m_\pi^2 g_{\pi NN}}{m_\pi^2-t}+
  \frac1{\pi} \int_{(3m_\pi)^2}^\infty ds \frac{{\rm Im} \hat m F_P(s)}{s-t} \, , 
\end{eqnarray}  
where $2 i {\rm Im} F_P(s)= F_P(s+i 0^+)-F_P(s-i 0^+) = {\rm Disc}
F_P(s)$ is the discontinuity across the branch cut. Condition 5 implies 
\begin{eqnarray}
2 M_N g_A &=& 2 f_\pi g_{\pi NN}+ \frac1{\pi}\int_{(3
  m_\pi)^2}^{\infty} ds \ \frac{\operatorname{Im}\hat{m}F_P(s)}{s} \, .
\label{eq:GT-corr}
\end{eqnarray}
The residue corresponds to the $g_{\pi NN}$ definition. 
Note that the variables appearing here are the {\it physical} ones, and
that the GT relation is obtained when the continuum states are
neglected. This is unlike the chiral limit, $m_\pi \to 0$, where the
GT involves {\it unphysical} quantities, $\overset{\circ}{M_N}
\overset{\circ}{g}_A = \overset{\circ}{f_\pi} \overset{\circ}{g}_{\pi
  NN} $
where $ \overset{\circ}{A}= A |_{m_\pi \to 0}$ and the discrepancy is obtained
by using ChPT.


The asymptotic conditions are equivalent to the following sum rules
(analogous to the Weinberg sum rules)
\begin{eqnarray}
0&=&  2 f_\pi m_\pi^2 g_{\pi NN} + \frac1{\pi}\int_{(3 m_\pi)^2}^{\infty} ds \ \operatorname{Im}\hat{m}F_P(s), \label{eq:cond1} \\ 
0&=&  2 f_\pi m_\pi^4 g_{\pi NN} + \frac1{\pi} \int_{(3 m_\pi)^2}^{\infty} ds \ \operatorname{Im}\hat{m}F_P(s) s \, . \label{eq:cond2}
\end{eqnarray}
In our case we are interested in the process $N \bar N \to X= \pi,
3\pi, 5\pi, \dots$ in the channel with pion quantum numbers.
\section{Anatomy of Spectral function}

As we see, phenomenologically $\Delta_{\rm GT} > 0$ and the three
integrals involving the the spectral funcion must be
negative. However, introducing the spectral function $\rho(s)= {\rm
  Im} \hat m F_P(s)/\pi$ for short and taking
\begin{eqnarray}
  \rho(s) = \left[ {\rm sign} (\rho(s)) \sqrt{|\rho(s)/s|} \right] \sqrt{|\rho(s)s|} \equiv f(s) g(s)\, , 
\end{eqnarray}
owing to Schwartz's inequality we get ,
\begin{eqnarray}
  \Big|\int ds \rho(s) \Big|^2 &=& \Big|\int ds f(s) g(s) \Big|^2 \nonumber  \\ 
  \le \int ds |f(s)|^2 \int ds |g(s)|^2  &=& \int ds |\rho(s)| s \int ds |\rho(s)|/s   \, . 
\end{eqnarray}
For a spectral function with a well defined sign one has either $
|\rho(s)|= \rho(s) $ or $
|\rho(s)|= -\rho(s) $
and so we can intertwine the three sum rules
yielding
\begin{eqnarray}
  \left( 2 f_\pi m_\pi^2 g_{\pi NN} \right)^2 &\le& \left( 2 f_\pi m_\pi^4 g_{\pi NN} \right) 
  \left( 2 f_\pi g_{\pi NN} - 2M_N g_A \right) \nonumber \\
  &\implies&  1 \le 1-\Delta_{\rm GT} \implies \Delta_{\rm GT} \le 0 \, ,  
\end{eqnarray}  
in contradiction with the phenomenological positive sign, $\Delta_{\rm GT}=2(1) \% $. Thus, the spectral
function must change sign. At threshold the sign is well defined, so
there must be {\it at least} one value where $\rho(s_0)=0$. The
presence of this zero in the spectral function underlies the relative
insensitivity of results for $g_{\pi NN}$ found in Ref.~\cite{ERA-PSP:2023} and
reviewed below. 

The dispersive integral in Eq.~(\ref{eq:GT-corr}) runs from the three-pion
threshold $\sqrt{s}= 3 m_\pi$ to infinity and our estimate will be
done by separating the contributions into three different regions
\begin{enumerate}
\item Low energy region, $3 m_\pi \le \sqrt{s} \le \Lambda_{\chi}$,  where we may rely on Chiral Perturbation Theory~\cite{Kaiser:2003dr,Kaiser:2019irl}. 
\item Intermediate energy region, $\Lambda_\chi \le \sqrt{s} \le
  \Lambda_{\rm H}$, where pseudoescalar ,$J^{PC}=0^{-+}$ , resonances
  contribute explicitly on top of an unknown but hopefully smooth
  background.
\item High energy region where we  can use Perturbative QCD (pQCD)~\cite{Alvegard:1979ui,Brodsky:1980sx}.
\end{enumerate}
The precise values marking the boundaries between the different
regions are not precisely known and varying their values provides a
natural source of uncertainty. It turns out that both the low- and high-energy 
regions are numerically supressed even for the extreme values
$\Lambda_\chi \sim \Lambda_{\rm H} \sim 1 {\rm GeV}$. 
Furthermore, since their relative contribution as well as their 
uncertainties are also negligible, we shall focus in these proceedings
on the intermediate resonance region.

In the $I^G J^{PC}=1^- 0^{-+}$ channel there are 5 states reported by
the PDG booklet~\cite{ParticleDataGroup:2022pth}, $\pi(140),\pi(1300),\pi(1800),\pi(2070),\pi(2360)$
which we will denote as $\pi,\pi'\pi'',\pi''', \dots$. The first three
states are firmly established while, the last two ones fall under the label
further states, meaning that they have been observed by some
experiments but not confirmed by further experiments. These
pseudoscalar states follow a nice and clear radial Regge pattern 
\begin{eqnarray}
  M_n^2 \pm \Gamma_n M_n = 1.27(27) n + M_1^2 \quad ({\rm GeV}^2) \qquad n\neq 0
\end{eqnarray}
 which has been analyzed in Ref~.\cite{Masjuan:2012gc} in terms of
 the half-width rule which takes into account the fact that, when
 produced, resonances are not pure mass states. They correspond to a 
 mass distribution and different processes have different backgrounds,
 so that we expect the resonance mass spread is comparable to the
 width, so that $\Delta M^2 = M \Gamma$ or $\Delta M= \Gamma/2$.
 Useful uncertainties estimates arise from this simple half-width rule~\cite{Masjuan:2012sk,RuizArriola:2012ius}. 

Within the (relativized) quark model of Isgur and
Godfrey~\cite{Godfrey:1985xj} these states are interpreted as excited
$\bar q q$ configurations where $n$ indicates the number of nodes of
the relative wave function in two body Hamiltonian which up to
hyperfine splitting corrections is of the form $H=2 \sqrt{p^2+m_q^2}+
\sigma r$ (here $m_q$ represents the constituent quark mass).

Experimentally these states are inferred from decay modes
characterized by masses and widths, say $\pi(1300) \to \rho \pi \to 3
\pi $ or $ \pi(1800) \to \sigma \pi \to 3 \pi $ in several production
processes from a partial wave analysis. The determination of resonance
profiles including relativistic centrifugal barrier effects~\cite{VonHippel:1972fg,Chung:1995dx} relies strongly on particular parameterizations which ultimately are validated by
fitting actual data. Thus, up to a (hopefully) smooth background one has 
\begin{eqnarray}
  \operatorname{Im}\hat{m}F_P(s) = \sum_{P=\pi',\pi'', \dots} 2 f_P g_{P NN} m_P^2 F(s,m_P^2) 
\, , 
\end{eqnarray}
where $F $ is some profile function containing resonance parameters and
kinematical factors fixed partly, but not entirely, by theoretical
constraints. In fairness, this is a model dependent ansatz which we
have shown~\cite{ERA-PSP:2023} on the light of several resonance models 
(e.g., different resonance profiles) to have
little impact on the final determination of $g_{\pi NN}$ or
$\Delta_{\rm GT}$. Thus, in our work we propose a rather simple
short-cut based on the large $N_c$ limit in QCD.

\section{Minimal hadronic ansatz}


In the large $N_c$ limit, resonances become narrow, and actually $\Gamma/m = {\cal}( 1/N_c)$ (numerically one has $\Gamma/m=0.12(8)$~\cite{RuizArriola:2011nrw,Masjuan:2012gc}), so that in that limit 
we expect 
\begin{eqnarray}
\operatorname{Im}\hat{m}F_P(s) = \sum_{P=\pi',\pi'', \dots} 2 f_P g_{P NN} m_P^2 \delta (s
- m_P^2) \, , 
\end{eqnarray}
for $ s \ge (3 m_\pi)^2 $,  
where for the excited pions we introduce their corresponding weak
decay constant $f_P$ (note that such decay constants must 
vanish as $m_q\to0$, that ultimately guarantees the correct chiral 
extrapolation in our model)  corresponding to the process $P \to \nu_\mu \mu
$, and their coupling to the nucleon, $g_{PNN}$. In this limit 
the result becomes model independent and this spectral function yields
a sum of monopoles for the pseudoscalar form factor which in the space-like region, ($s=-Q^2 < 0$),  reads 
\begin{eqnarray}
  \hat{m}F_P(s=-Q^2) = \sum_{P=\pi, \pi',\pi'', \dots}  \frac{ 2 f_P g_{P NN}  m_P^2}{Q^2+ m_P^2} \, , 
\end{eqnarray}
which can be reinterpreted as an Extended PCAC relation and corresponds to flat strong $PNN$ vertices. The minimal hadronic ansatz fulfilling all conditions 
involving $\pi$ as well as
the lowest excitations, $\pi'$ and $\pi''$ yields 
\begin{eqnarray}
F_P (t)= 2 M g_A \frac{1}{1-t/m_\pi^2} \frac{1}{1-t/m_{\pi'}^2}  \frac{1}{1-t/m_{\pi''}^2}  
\end{eqnarray}
which only requires PDG masses. From the residue at the pion pole we get 
\begin{eqnarray}
g_{\pi NN}= \frac{M_N g_A}{f_\pi} \frac{1}{1-m_\pi^2/m_\pi'^2} \frac{1}{1-m_\pi^2/m_\pi''^2}  \, , 
\end{eqnarray}
so that the GT discrepancy becomes
\begin{eqnarray}
  \Delta_{\rm GT}=1-\frac{M g_A}{g_{\pi NN} f_\pi} = m_\pi^2 \left[ \frac{1}{m_\pi'^2} + \frac{1}{m_\pi''^2} \right]  -  \frac{m_\pi^4}{m_\pi'^2 m_\pi''^2}  \, , 
\end{eqnarray}
which is insensitive to $M, g_A, f_{\pi}$.
The recommented PDG values are $m_{\pi'}=1.3(1)$GeV and
$m_{\pi''}=1.81(1)$GeV. Using these values yields $\Delta_{\rm GT}=
1.7(2)\%$. If, instead we average the PDG listed most likely values
for any experiment for the $\pi'$ and $\pi''$ masses, we get
$\Delta_{\rm GT}= 2.3(3)\%$. These values are in the bulk with the
current estimate. A more conservative estimate is based
on using the half-width rule, so that from $\Gamma_{\pi'}=0.3(1)$GeV and
$\Gamma_{\pi''}=0.22(3)$GeV one gets $\Delta_{\rm GT}= 1.7(4)\%$.  In the
worst possible scenario of the recommended PDG values,
$\Gamma_{\pi'}=0.6$GeV , one obtains $\Delta_{\rm GT}= 2(1)\%$. 
We note that the values here provided are yet 
provisional ones.

\section{Conclusions}

While these findings are by themselves rather satisfactory, it is
worth to note that they efficiently summarise a more complete
treatment incorporating the role of chiral corrections at low
energies, finite width effects, modifications due to excited states
beyond the $\pi(1800)$ and high energy perturbative QCD
contributions~\cite{ERA-PSP:2023}.  Besides its simplicity this
phenomenological estimate is  competitive with much
more involved current analyses requiring a large nucleon-nucleon
scattering database below pion production threshold.
This also opens up the possibility to study similar 
cases, such as the octet and singlet currents, which current 
knowledge is less advanced than the isovector case here discussed.

\end{document}